\documentclass[fleqn, usenatbib]{mnras}
\usepackage[T1]{fontenc}
\usepackage{ae,aecompl}
\usepackage{graphicx}
\usepackage{amsmath}	
\usepackage{amssymb}	
\graphicspath{{./}{figures/}}

\title[KIC~9821622]{ Concerning Li-rich status of KIC~9821622: A Kepler field RGB star reported as Li-rich Giant}

\author[Singh et al. 2019]
{
 Raghubar Singh,$^{1,2}$ \thanks{E-mail:raghubar.singh@iiap.res.in}
 Yerra Bharat Kumar,$^{3}$
 Bacham E. Reddy,$^{1, 3}$
 Wako Aoki $^{4}$ \\
 $^{1}$ Indian Institute of Astrophysics, II Block, Koramangala, Bengaluru-560034, India\\
 $^{2}$ Pondicherry University R. V. Nagara, Kala Pet, 605014, Puducherry, India\\
 $^{3}$ Key Laboratory of Optical Astronomy, National Astronomical Observatories, Chinese Academy of Sciences, Beijing, 100101, China \\
 $^{4}$ National Astronomical Observatory of Japan, Mitaka, Tokyo 181-8588, Japan 
}
\date{Accepted XXX. Received YYY; in original form ZZZ}

\pubyear{2019}

\begin{document}
\label{firstpage}
\pagerange{\pageref{firstpage}--\pageref{lastpage}}
\maketitle

\begin{abstract}
 
 Given the implications for the origin of Li enhancement in red giants we have reviewed Li-rich classification of KIC~9821622, the only bonafide RGB giant with He inert-core till date,  reported as a Li-rich giant by reanalyzing the high-resolution spectra. We have obtained $A(Li)_{LTE} = 1.42 \pm 0.05$ dex. After correcting for non-LTE it is $A(Li)_{NLTE} = 1.57 \pm 0.05 $ dex which is significantly less than the reported A(Li) = $1.80 \pm 0.2$~dex. We found the sub-ordinate line at 6103 \AA\ is too weak or absent to measure Li abundance. The derived abundance is normal for red giants undergoing dilution during the 1st dredge-up. Since all the known Kepler field Li-rich giants belong to the red clump region, this clarification removes the anomaly and strengthens the evidence that the Li enhancement in low mass giants may be associated only with the He-core burning phase. The Li excess  origin, probably, lies during He-flash at the RGB tip, an immediate preceding phase to red clump. 
 
\end{abstract}


\section{Introduction} 

 It is now accepted that there is a small group of red giants with significantly higher photospheric Li abundances compared to maximum values of A(Li) = 1.6 to 1.8~dex predicted by standard models \citep{Iben1967a, Iben1967b, Lagarde2012}. In some cases, Li abundances in these giants exceed by a factor of 10 to 1000 compared to predictions \citep{Brown1989, Charbonnel2000, Kumar2011a, yan2018, deepak2019}. However, the origin of high Li abundance in these giants has been a subject of debate for over three decades since their first discovery in 1982 \citep{Wallerstein1982}. This is mainly because of a lack of precise determination of Li-rich giants' exact location on RGB \citep[see e.g;][]{Charbonnel2000}. In particular, uncertainty in determining whether a particular Li-rich giant is at RGB bump or at red clump as their positions overlap in $T_{\rm eff}-L$ plane \citep[e.g;][]{Kumar2011a}. This uncertainty led to multiple theories \citep[e.g.;][]{Delareza1996, Siess1999, Palacios2001, Denissenkov2004, Denissenkov2012}. 
 
 Now it is possible, thanks to Kepler \citep{Borucki2010} asteroseimic data, to separate giants with inert He-core ascending RGB for the first time from those of red clump giants of He-core burning \citep{Bedding2011}. A recent study by \cite{Singh2019l} showed that all Li-rich giants for which asteroseismic analysis is available are indeed red clump giants in the He-core burning phase. This is quite significant as this survey is based on large unbiased data set drawn from Kepler \citep{Borucki2010} and LAMOST \citep{Cui2012} catalogs. This study reconfirms earlier serendipitous discoveries of five Li-rich giants, all of them have been classified as red clump giants based on asteroseismic data \citep{SilvaAguirre2014, Carlberg2015, kumar2018, Smiljanic2018}. The mounting evidence that Li-rich phenomenon is probably associated only with He-core burning giants, post-He-flash, has been further strengthened by a recent large survey of Li-rich giants among GALAH \citep{galah2015} and GAIA \citep{Gaiadr2} catalogs by \cite{deepak2019}, in which they showed a vast majority of Li-rich giants belong to red clump region.  

  KIC~9821622 is the lone exception among Li-rich giants for which stellar evolutionary phase is determined using robust asteroseismic technique. Exception being, it is the only bonafide red giant with inert He-core reported as Li-rich giant \citep{Jofre2015} and all the others are associated with core He-burning phase of red clump, post He flash. It is interesting to see whether this star is indeed a Li-rich giant as its Li abundance is very close to the border. Understanding of its Li-rich status has implications for Li origin scenarios. If this is indeed a Li-rich giant as reported, one may have to revoke multiple theories for Li enrichment among red giants such as planet engulfment for occurrence of Li-rich giants anywhere along the RGB \citep[e.g.;][]{Siess1999} or extra mixing during bump luminosity, and He-flash combined with non-canonical mixing for Li-rich RC giants, etc. In this paper, we present clarification based on results obtained using high-resolution spectra taken from the archives of Gemini HARPS and Subaru HDS. 
 
 \begin{table}
 \caption{Stellar parameters and Li abundance of KIC~9821622 from 4 different studies.}
 \begin{tabular}{l|l|l|l|r}
 \hline
 Parameter & J15 & T17 & Y16  & This work\\[0.25em]
 \hline
 $T_{\rm eff}$ (K)  & $4725 \pm 20 $ & 4896 & 4895 & 4750$\pm$30 \\ [0.25em]
 $\mathrm{[Fe/H]}$ & $ -0.49 \pm 0.03$ &-0.25 & -0.40 & -0.49$\pm$0.06 \\ [0.25em]
 $\Xi_{t}$ (Km s$^{-1}$) & $ 1.12 \pm 0.04$ & 1.03 & 1.18 & 1.08$\pm$0.1 \\ [0.25em]
 log~g & $ 2.71 \pm 0.09$ &2.91 & 2.71 & 2.73$\pm$0.1 \\ [0.25em]
 vsini (Km s$^{-1}$) & $1.01\pm 0.77  $ & 1.9 & --- & 1.9$\pm$0.5 \\ [0.25em]
 A(Li)6103 & $1.80 \pm 0.04$ & 1.67 & --- & --- \\ [0.25em]
 A(Li)6707 & $1.49 \pm 0.02$& 1.85 & 1.63 & 1.42$\pm$0.05 \\ [0.25em]
 \hline
 \label{tab:tab1}
 \end{tabular}
 \vskip -0.1cm
 {J15: \cite{Jofre2015}; T17: \cite{Takeda2017}; Y16: \cite{Yong2016}}
  \end{table}

\section{Data acquisition}

 For this study, we have used wavelength calibrated and continuum normalized spectra taken using HARPS spectrograph equipped to Gemini telescope \url{https://www.gemini.edu/sciops/instruments/graces/2015-onsky-tests}. Spectra has spectral resolution of R$\approx$67500, wavelength range of 3990 -10479 \AA, and S/N $\approx$ 150 for median combined spectra. We also used reduced spectra from Subaru archive \url{https://smoka.nao.ac.jp/} with spectral resolution of R$\approx$80000 and S/N $\approx$ 50. 

 \begin{figure}
    \centering
    \includegraphics[width=0.50\textwidth]{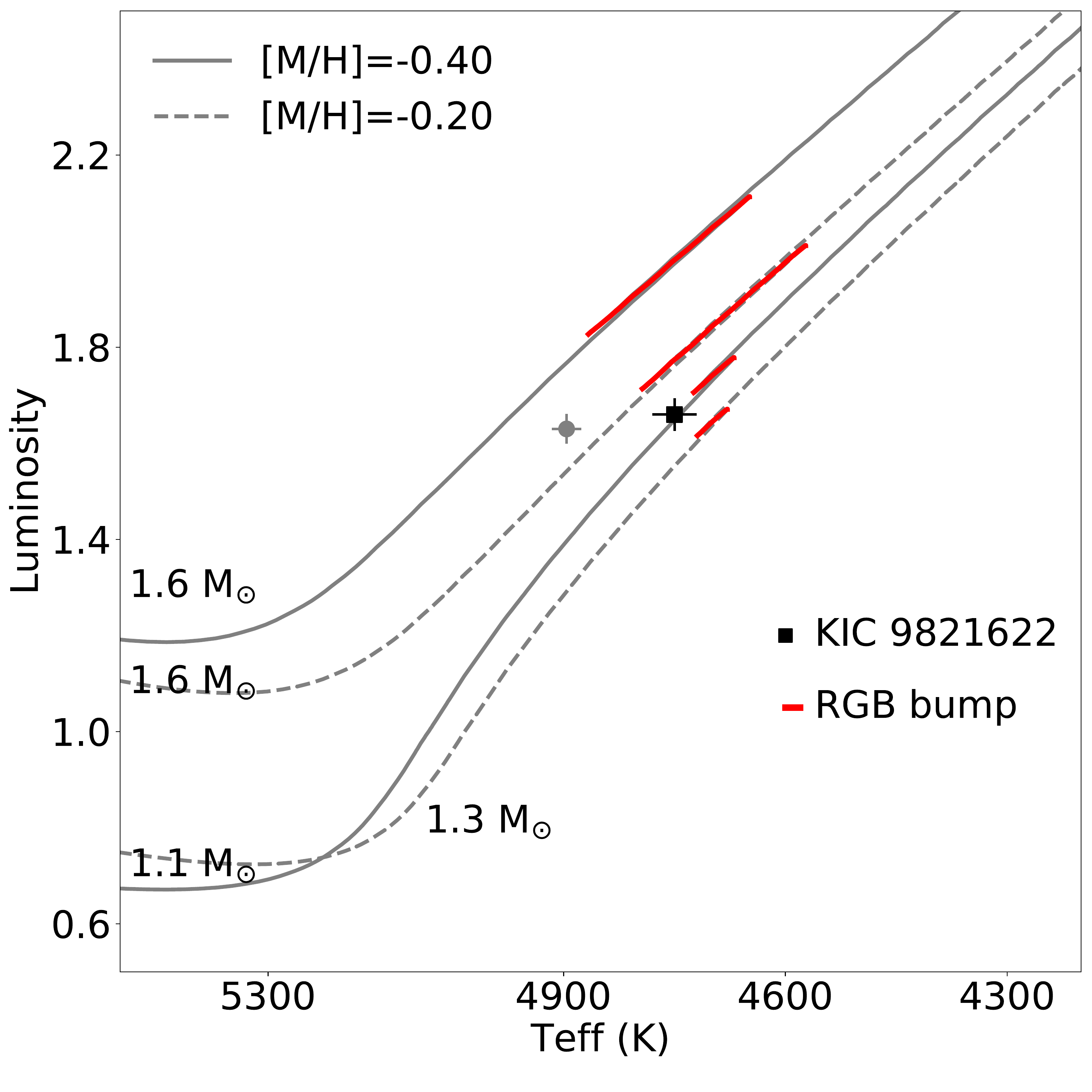}
    \caption{KIC~9821622 in HR-diagram based on derived stellar parameters. Tracks are from Parsec models. Dotted tracks are scaled up evolutionary tracks for [$\alpha$/Fe] = 0.32 for the same derived mass of 1.6~M$_{\odot}$ (see text for more details). Note, its position below the luminosity bump.}
    \label{fig:bhrd}
 \end{figure}

 \begin{figure}
    \centering
    \includegraphics[width=0.49\textwidth]{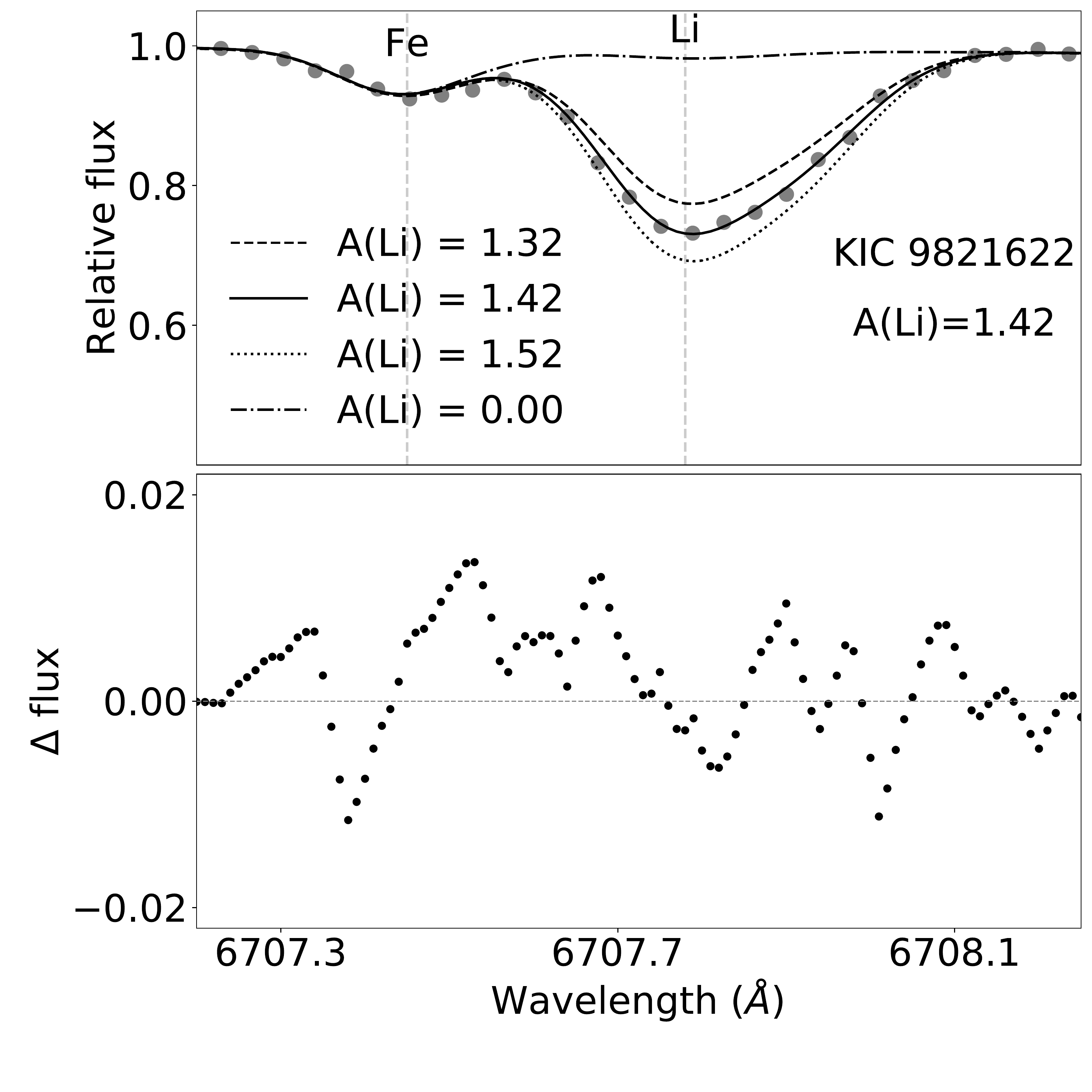}
    \caption{ Comparison of observed HARPS and synthetic spectra at Li resonance line  at 6707.78 \AA\ for different Li abundances. Synthetic spectra (black solid line) for  A(Li)=1.42 dex best fits observed line. In Bottom panel difference between observed and best fit synthetic spectra is shown.}
    \label{fig:lisyn}
 \end{figure}

 \begin{figure}
    \centering
    \includegraphics[width=0.50\textwidth]{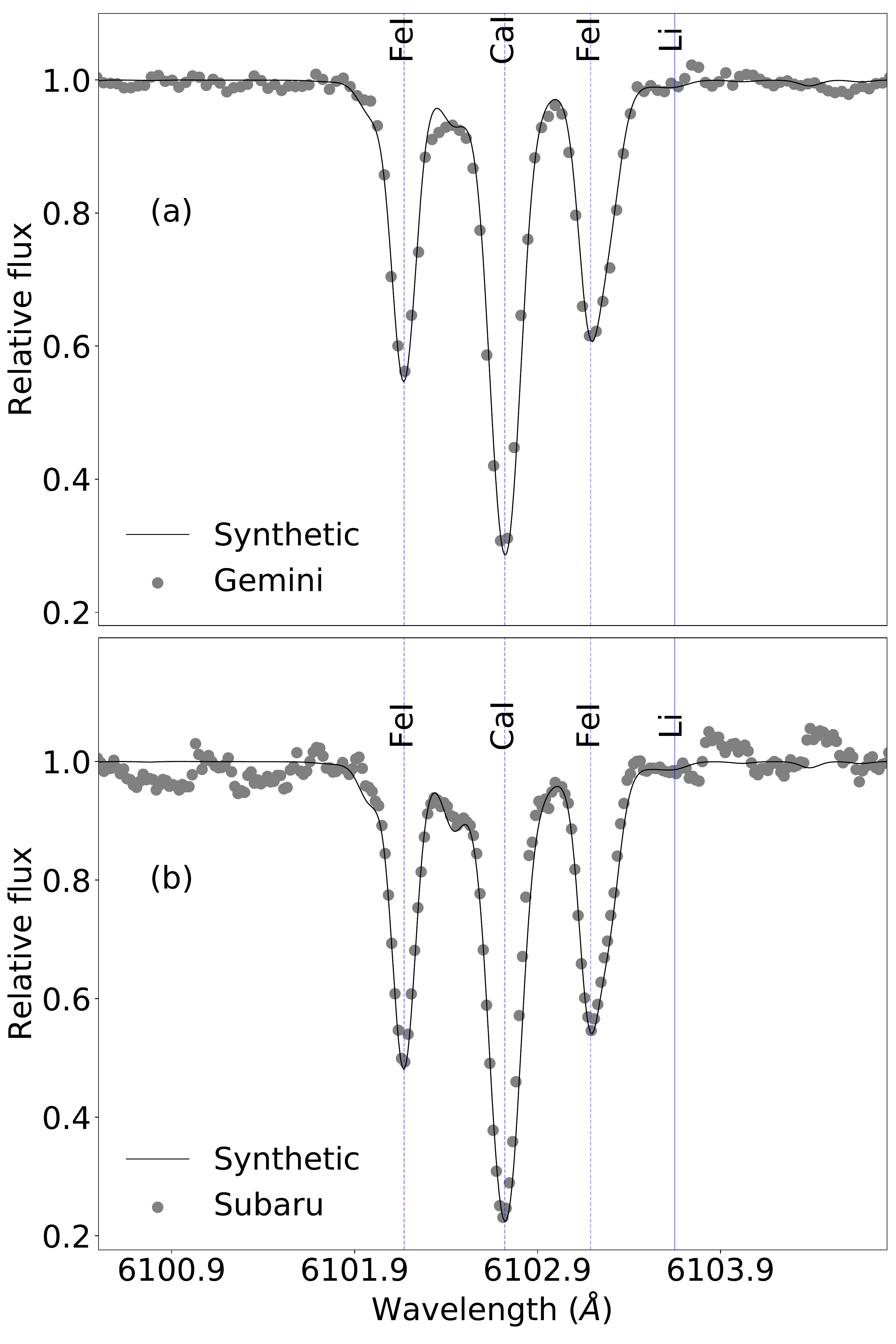}
    \caption{Synthesis of spectra in the region of 6100--6105 \AA. Li line is clearly not visible because of relatively low SNR. }
    \label{fig:lisyn2}
 \end{figure}

\section{Atmospheric Parameters and Li abundance}

\begin{table}
\caption{Stellar parameters of KIC~9821622 based on spectroscopy and asteroseismology.}
 \begin{tabular}{l|c|r|}
 \hline
 Star & KIC 9821622 & Remarks\\
 \hline
 \hline
 RA & 19:08:36.16 & \\
 Dec & 46:41:21.25 & \\ 
 Vmag & $12.22 \pm 0.04 $ & APASS\\
 Parallax (mas) & $0.556 \pm 0.02$ & Gaia DR2\\
 BC & $-0.35 \pm 0.034$ & \\
 $ \log(L/L_{\odot})$ & $ 1.66 \pm 0.034$ & \\
 $ \Delta p$ (sec) & $67.6 \pm 1.6$ & \cite{Mosser2014} \\
 $ \Delta \nu$ & $ 6.02 \pm 0.015$ & \cite{Yu2018} \\
 $ \nu_{max}$ & $ 63.39 \pm 0.57 $ & \cite{Yu2018} \\
%  Radius (R$_{\odot}$) & $9.37 \pm 0.16$ & \\
 log~g$_{seis}$ & $2.71 \pm 0.03$ & \\
 log~g$_{Gaia}$ & 2.63 & \\
 Vmacro & $3.7$ & \\
 A(Li)$_{LTE}$ & $1.42 \pm 0.05$ & \\
 A(Li)$_{NLTE}$ & $1.57 \pm 0.05$ & \\
 $\mathrm{[Eu/Ba]}$ & $0.53 \pm 0.1$& \\
 $\mathrm{[C/N]}$  & $0.20 \pm 0.02$& \\
 N/O & 0.15 & \\
%  Age (Gyrs) & $ 2.25 \pm 0.3$ \\
 \hline
 Age (Gyr) & 2.482 $\pm$ 0.164  & PARAM \\
 Mass (M$_{\odot}$) & 1.465 $\pm$ 0.033 & PARAM \\
 R (R$_{\odot}$) & 8.942 $\pm$ 0.095 & PARAM \\
 log~g & 2.693 $\pm$ 0.003  & PARAM\\
 \hline
 Photometry &  & \\ [0.2 em]
 \hline
 (B-V) &    $1.012 \pm 0.04$ &  \\
 Teff (B-V) & $4665 \pm 60$  & \\
 (V-K) &  $2.378 \pm 0.04$ & \\
 Teff (V-K) & $4730 \pm 40$  & \\
 \hline
 APOGEE &  &  \\ [0.2 em]
 \hline
 Teff & $4744 \pm 69$ &  \\
 log~g & $2.493 \pm 0.08$ & \\
 $\mathrm{[Fe/H]}$ & $-0.358 \pm 0.029$ & \\
 \hline
 \label{tab:tab2}
 \end{tabular}
  
 \end{table}

 Stellar atmospheric parameters ($T_{\rm eff}$, log $g$, and [Fe/H]) have been derived using standard method based on LTE stellar atmospheric models \citep{CastelliKuruxz2004} and MOOG \citep{Sneden1973}. For this, a good number of \ion{Fe}{i} and \ion{Fe}{ii} with well determined oscillator strengths and relatively free of blends are taken from \cite{Reddy2003, Ramirez2011, Carlberg2012, Liu2014}. $T_{\rm eff}$ has been derived through iterative process in which derived \ion{Fe}{i} abundance is independent of line low excitation potential for a set of given log $g$, microturbulance ($\xi_{t}$) and [Fe/H]. Similarly, $\xi_{t}$ has been derived for which \ion{Fe}{i} abundance is independent of line equivalent widths, and log $g$ is the one for which derived abundances of \ion{Fe}{i} and \ion{Fe}{ii} are equal. Uncertainties in $T_{\rm eff}$ and $\xi_{t}$ are evaluated based on sensitivity of the slope of relation between abundances and LEP, and abundances and EWs to changes in respective parameters. The derived values are given in Table~\ref{tab:tab1}. $T_{\rm eff}$ is also derived from photometry using (V-K) calibration from \cite{Alonso1999} (See Table~\ref{tab:tab2}). Our derived values are similar to the values derived by \cite{Jofre2015} and infrared APOGEE spectra \citep{Majewski2017, Garcia2016}. However, these values are cooler by about 140~K compared to two other recent studies by \cite{Takeda2017, Yong2016} (See Table~\ref{tab:tab1}). The stellar age along with mass, radius, and log$g$ are computed from asteroseismic parameters ($\nu _{max}$, $\delta \nu$) and derived spectroscopic parameters ([Fe/H], $T_{\rm eff}$) using PARAM based on Bayesian statistics\footnote{\url{http://stev.oapd.inaf.it/cgi-bin/param_1.3}} code \citep{desilva2006}. Derived values are given in Table~\ref{tab:tab2}. However, the calibration relation \citep{Bedding1995} for the same input asteroseismic parameters and derived $T_{\rm eff}$ returns mass of $1.62 \pm 0.05$~M$_{\odot}$. Note, there is no [Fe/H] parameter in this relation.
 
 Add to the complication, the adopted $T_{\rm eff}$ = 4750~K and luminosity derived using Gaia astrometry combined with PARSEC evolutionary tracks \citep{Bressan2012} as shown in Figure~\ref{fig:bhrd} suggest stellar mass is close to 1.1~M$_{\odot}$ which is significantly less than the mass = 1.62~M$_{\odot}$ derived from asteroseismology. However, higher $T_{\rm eff}$ = 4890~K derived by \cite{Yong2016, Takeda2017} yield  mass which is close to the value derived from asteroseismology. In either case, star's position is below the luminosity bump (See Figure~\ref{fig:bhrd}). We tested whether the input value of $T_{\rm eff}$ in deriving mass using asteroseismic parameters plays a role. We find little difference in mass ($\approx$ 0.1~M$_{\odot}$) from the two values of $T_{\rm eff}$. Since the adopted $T_{\rm eff}$ in this study seems to be robust as this agrees well with the least extinction affected (V-K) colour temperature and values from the APOGEE infrared spectra (see Table~\ref{tab:tab2}), the discrepancy in $T_{\rm eff}$ and, thereby, in its mass is, probably, due to giant's peculiar nature. KIC~9821622 is a known member of a small group of young $\alpha$- and $r$-process enriched giants \citep{Matsuno2018}. It has [$\alpha$/Fe] = 0.32~dex, which is quite high for its [Fe/H] = $-$0.49~dex \citep{Reddy2003}. If we use $\alpha$-enhanced evolutionary tracks instead of normal ones, the discrepancy in mass  will significantly reduce as $\alpha$ enhanced tracks of same mass are relatively cooler compared to normal ones \citep[see,][]{Fu2018}. A scaled-up model based on \cite{Fu2018} study will be metal-rich by about 0.3~dex for [$\alpha$/Fe] = 0.32~dex enhanced models. The scaled model of 1.6~M${_\odot}$ is shown as dotted lines in Figure~\ref{fig:bhrd}. As shown, the discrepancy in mass derived from asteroseismic analysis and $\alpha$-element enhanced evolutionary tracks is small about 0.15~M$_{\odot}$. PARAM returns for the same parameters with scaled-up [Fe/H] = -0.2~dex, a mass of 1.51~M$_{\odot}$. This amounts to a gap of 40~K to 50~K in $T_{\rm eff}$ between the two extreme masses. Also, note among various RGB evolutionary tracks there is an uncertainty of about 50 - 60~K \citep[see e.g.;][]{Tayar2017, Choi2018}. 
 
 Li-abundance is derived from Li-resonance line at 6707.78 \AA\ using HARPS spectrum of S/N $\approx$ 150 by comparing synthetic spectra to observed one (see Figure~\ref{fig:lisyn}). Synthetic spectra is generated using Kurucz model atmosphere \citep{CastelliKuruxz2004} and LTE radiative transfer code MOOG \footnote{\url{https://www.as.utexas.edu/~chris/moog.html}}, using 'synth' driver \citep{Sneden1973}. For spectral synthesis, we used line list compiled by \cite{Reddy2002} for a resonance line at 6707 \AA\ which includes hyperfine structure and molecular lines. We also examined spectra synthesized using the same line list used by \citep{Jofre2015} and find no significant difference between the two line lists. Synthetic spectra computed for A(Li)= 1.42~dex well matches with the observed line profile with an uncertainty of 0.05~dex which is a cumulative uncertainty estimated by quadratic sum of uncertainties in derived stellar parameters. Our derived abundance is very similar to the value A(Li) = 1.49~dex obtained by \cite{Jofre2015} from the same Li line. However, our value is about 0.2~dex lower than the value derived by \cite{Yong2016} which is expected as their $T_{\rm eff}$ is higher by about 150~K. Both \cite{Jofre2015} and \cite{Yong2016} used Gemini HARPS spectra. On the other hand \cite{Takeda2017} used Subaru spectra of S/N $\approx$ 50. They report significantly higher A(Li) = 1.85~dex. Derived Li abundance from all the three studies along with ours are given in Table~\ref{tab:tab1}. Apart from the resonance line at 6707 \AA\ there is another Li transition at 6103 \AA\ which is weak, generally seen in stars with large Li abundances. In Figure~\ref{fig:lisyn2}, we displayed spectra of KIC~9821622 at 6103 \AA\ region taken from both Gemini HARPS and Subaru. Li transition at 6103 \AA\ line is not seen in either of the spectra. The reported A(Li) = 1.80 ~dex from this line by \cite{Jofre2015} appears to be an upper limit. \cite{Yong2016} performed abundance analysis but didn't report Li abundance from 6103 \AA\ line.  Given the well defined Li transition at 6707 \AA\ and its reliability for A(Li) measurement, and absence of Li transition at 6103 \AA, we don't consider Li abundance reported from  6103 \AA\ line any further.   

\section {discussion}

 What is the expected photospheric Li abundance in RGB stars? According to standard models \citep{Iben1967a} Li abundance is a function of stellar mass and its initial Li abundance with which stars evolved off. Also, to certain extent, it is  metallicity dependent. Radiative opacity is relatively lower in lower metallicity stars, and hence less efficient mixing during their evolution on RGB compared to metal-rich stars of same metallicity \citep{Lagarde2012}. As a result, rate of photospheric Li abundance depletion is lower in low metallicity giants of same mass. Further, standard models predict depletion of Li abundance only during 1st dredge-up, and expected no further dilution as the convection envelope recedes from the hydrogen burning shell. Standard models \citep{Iben1967a} set 1st dredge-up depletion  upper limits of about A(Li) = 1.6 to 1.8~dex for Li normal giants of mass between 1.0 to 1.5~M$_{\odot}$. 
 
 However, dilution of Li post 1st dredge-up seems to happen as giants evolve with luminosity via bump as neatly illustrated by observations of giants in globular cluster NGC~6397 \citep{Lind2009b} and predicted by theoretical models of extra mixing (e.g. \citealt{Lagarde2012}). Thus, apart from mass and metallicity, it is important to consider giants' evolutionary phase as well to determine whether Li abundance is normal or enhanced in any particular giant \citep{Kirby2016}. It's position in HR diagram (Figure~\ref{fig:bhrd}) coupled with the derived values of $^{12}C/^{13}C$ and [C/N] ratios (see Table~\ref{tab:tab2}) confirms that the star is not evolved past the luminosity bump.
 
 \begin{figure*}
   \centering
   \includegraphics[width=0.99\textwidth]{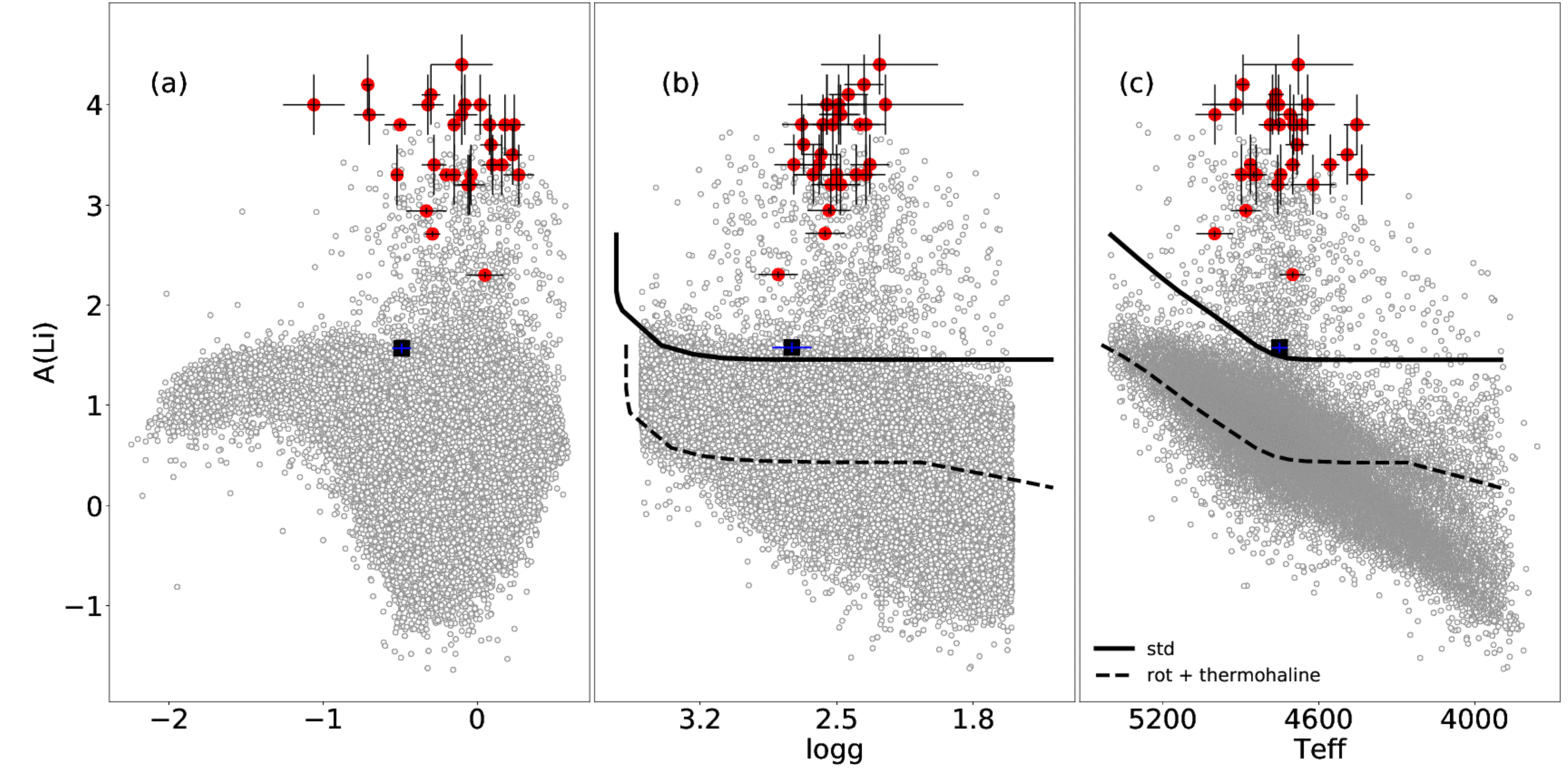}
   \caption{Location of KIC~9821622 in A(Li) vs [Fe/H], Vs log$g$, and Vs $T_{\rm eff}$ plots. Background stars are from the catalogue of GALAH survey. Sample giants are low mass (M $\leq$ 2~M$_{\odot}$) red giants with Flag$_{canon}$ = 0 for atmospheric parameters (see \citealt{deepak2019}). Solid and dashed lines are models of 1.5~M$_{\odot}$ from \citealt{Lagarde2012} for standard mixing and thermohaline mixing with rotation included, respectively. In all the panels red filled circles are the known Li-rich giants classified as red clump giants using asteroseismic analysis.}
   \label{fig:lagli}
 \end{figure*}
 
 Let us examine model prediction of Li abundance upper limit for KIC~9821662 at its present evolutionary position for its stellar parameters and initial A(Li). We don't have exact Li abundance with which this particular star might have evolved off. However, we can use general trends of A(Li) with mass and [Fe/H] among main sequence dwarfs. We adopt initial A(Li) from a study of \cite{Lambert2004} who provides plots of A(Li) versus stellar mass for different bins of [Fe/H]. For metallicity of [Fe/H] $\approx$ -0.5 and $M\approx$ 1.5~M$_{\odot}$ we find maximum initial abundance in the range of A(Li) = 3.0 to 2.8~dex. Combining this information with \cite{Iben1967a} models (see their Fig. 21) we find Li deletion factor of about 12 for KIC~9821662 for its derived $T_{\rm eff}$ = 4750~K and for 1.5~M$_{\odot}$ model which correspond to values of about A(Li) = 1.92~dex, post 1st dredge-up. Even if we assume  lower initial values of A(Li) = 2.8~dex, we find A(Li) = 1.72~dex at the end of 1st dredge-up. We find even higher upper limits for higher values of $T_{\rm eff}$ = 4890~K as reported by \cite{Yong2016, Takeda2017}. For example, for the same 1.5~M$_{\odot}$,  models predict less Li depletion with final values corresponding to 2.0 and 1.89~dex for the respective initial values of Li abundance. This is in good agreement with the empirical relation between A(Li) and Mv derived based on cluster data by \cite{Kirby2016}. For its Mv = 0.99 (log (L/L$_{\odot}$) = 1.66) the relation suggests giant should have value more than A(Li) = 1.66~dex, in other words, the measured A(Li) = 1.57~dex for KIC~9821622 is normal for giants below the bump luminosity. 

 Observations of Li abundance in numerous studies \citep[e.g.][]{Brown1989} confirm upper limits of Li for normal giants predicted by classical models. This has been beautifully illustrated in a plot of A(Li) versus [Fe/H] for a large sample of giants taken from GALAH catalogue and shown by \cite{deepak2019} (See Figure~\ref{fig:lagli}). Within large [Fe/H] range, as defined by observations, upper limit is about 1.7~dex as measured by eye. Upper limits are low at the extreme ends of sample metallicity range. We won't comment on it as this is not a focus of present study. In the same figure, we have over plotted all the Li-rich giants (red symbols) for which asteroseismic classification is available including the star in question, KIC~9821622 (black square). Same data is shown as A(Li) vs log~$g$ (Figure~\ref{fig:lagli}~b) and A(Li) vs $T_{\rm eff}$ (Figure~\ref{fig:lagli}~c) along with the standard Li depletion models evolved from initial A(Li) of 3.3~dex \citep{Lagarde2012} for 1.5~M$_{\odot}$. Note, KIC~9821622 is the only giant with significantly less A(Li) which falls closer to the upper limit.
 
\section{Conclusion}

 We have performed a review of the reported Li-rich status of KIC9821622, a bonafide red giant ascending RGB. Re-analysis of the spectra based on high resolution archival spectra of Gemini HARPS and Subaru, and comparison with theoretical models suggest the giant is most probably a normal Li giant (A(Li) = $1.57 \pm 0.05$~dex) at the end of 1st dredge-up. The sub-ordinate line at 6103 \AA\ is very weak or absent for deriving Li-abundance contrary to previous report by \cite{Jofre2015}. Till date, all the Li-rich giants for which evolutionary phase is determined by a robust technique using asteroseismic analysis, are red clump giants with He-core burning phase.  This clarification will constrain theoretical models.

\section*{Acknowledgement}

 We thank anonymous referee for useful comments. 
 B.E.R thanks NAOC, Beijing for support through the CAS PIFI grant. 
 Y.B.K thanks NSFC for support through grant no. 11850410437.
 Based on observations obtained at the Gemini Observatory, which is operated by the Association of Universities for Research in Astronomy, Inc., under a cooperative agreement with the NSF on behalf of the Gemini partnership: the National Science Foundation (United States), National Research Council (Canada), CONICYT (Chile), Ministerio de Ciencia, Tecnolog\'{i}a e Innovaci\'{o}n Productiva (Argentina), Minist\'{e}rio da Ci\^{e}ncia, Tecnologia e Inova\c{c}\~{a}o (Brazil), and Korea Astronomy and Space Science Institute (Republic of Korea). This work has made use of data from the European Space Agency (ESA) mission {\it Gaia} (\url{https://www.cosmos.esa.int/gaia}), processed by the {\it Gaia} Data Processing and Analysis Consortium (DPAC, \url{https://www.cosmos.esa.int/web/gaia/dpac/consortium}). Funding for the DPAC has been provided by national institutions, in particular the institutions participating in the {\it Gaia} Multilateral Agreement. 

% \bibliographystyle{mnras}
% \bibliography{ref}

\begin{thebibliography}{}
\makeatletter
\relax
\def\mn@urlcharsother{\let\do\@makeother \do\$\do\&\do\#\do\^\do\_\do\%\do\~}
\def\mn@doi{\begingroup\mn@urlcharsother \@ifnextchar [ {\mn@doi@}
  {\mn@doi@[]}}
\def\mn@doi@[#1]#2{\def\@tempa{#1}\ifx\@tempa\@empty \href
  {http://dx.doi.org/#2} {doi:#2}\else \href {http://dx.doi.org/#2} {#1}\fi
  \endgroup}
\def\mn@eprint#1#2{\mn@eprint@#1:#2::\@nil}
\def\mn@eprint@arXiv#1{\href {http://arxiv.org/abs/#1} {{\tt arXiv:#1}}}
\def\mn@eprint@dblp#1{\href {http://dblp.uni-trier.de/rec/bibtex/#1.xml}
  {dblp:#1}}
\def\mn@eprint@#1:#2:#3:#4\@nil{\def\@tempa {#1}\def\@tempb {#2}\def\@tempc
  {#3}\ifx \@tempc \@empty \let \@tempc \@tempb \let \@tempb \@tempa \fi \ifx
  \@tempb \@empty \def\@tempb {arXiv}\fi \@ifundefined
  {mn@eprint@\@tempb}{\@tempb:\@tempc}{\expandafter \expandafter \csname
  mn@eprint@\@tempb\endcsname \expandafter{\@tempc}}}

\bibitem[\protect\citeauthoryear{{Alonso}, {Arribas}  \&
  {Mart{\'\i}nez-Roger}}{{Alonso} et~al.}{1999}]{Alonso1999}
{Alonso} A.,  {Arribas} S.,   {Mart{\'\i}nez-Roger} C.,  1999, \mn@doi [\aaps]
  {10.1051/aas:1999521}, \href
  {https://ui.adsabs.harvard.edu/abs/1999A&AS..140..261A} {140, 261}

\bibitem[\protect\citeauthoryear{{Bedding} et~al.,}{{Bedding}
  et~al.}{2011}]{Bedding2011}
{Bedding} T.~R.,  et~al., 2011, \mn@doi [\nat] {10.1038/nature09935}, \href
  {http://adsabs.harvard.edu/abs/2011Natur.471..608B} {471, 608}

\bibitem[\protect\citeauthoryear{{Bharat Kumar}, {Singh}, {Eswar Reddy}  \&
  {Zhao}}{{Bharat Kumar} et~al.}{2018}]{kumar2018}
{Bharat Kumar} Y.,  {Singh} R.,  {Eswar Reddy} B.,   {Zhao} G.,  2018, \mn@doi
  [\apjl] {10.3847/2041-8213/aac16f}, \href
  {http://adsabs.harvard.edu/abs/2018ApJ...858L..22B} {858, L22}

\bibitem[\protect\citeauthoryear{{Borucki} et~al.,}{{Borucki}
  et~al.}{2010}]{Borucki2010}
{Borucki} W.~J.,  et~al., 2010, \mn@doi [Science] {10.1126/science.1185402},
  \href {http://adsabs.harvard.edu/abs/2010Sci...327..977B} {327, 977}

\bibitem[\protect\citeauthoryear{{Bressan}, {Marigo}, {Girardi}, {Salasnich},
  {Dal Cero}, {Rubele}  \& {Nanni}}{{Bressan} et~al.}{2012}]{Bressan2012}
{Bressan} A.,  {Marigo} P.,  {Girardi} L.,  {Salasnich} B.,  {Dal Cero} C.,
  {Rubele} S.,   {Nanni} A.,  2012, \mn@doi [\mnras]
  {10.1111/j.1365-2966.2012.21948.x}, \href
  {https://ui.adsabs.harvard.edu/abs/2012MNRAS.427..127B} {427, 127}

\bibitem[\protect\citeauthoryear{{Brown}, {Sneden}, {Lambert}  \&
  {Dutchover}}{{Brown} et~al.}{1989}]{Brown1989}
{Brown} J.~A.,  {Sneden} C.,  {Lambert} D.~L.,   {Dutchover} Jr. E.,  1989,
  \mn@doi [\apjs] {10.1086/191375}, \href
  {http://adsabs.harvard.edu/abs/1989ApJS...71..293B} {71, 293}

\bibitem[\protect\citeauthoryear{{Carlberg}, {Cunha}, {Smith}  \&
  {Majewski}}{{Carlberg} et~al.}{2012}]{Carlberg2012}
{Carlberg} J.~K.,  {Cunha} K.,  {Smith} V.~V.,   {Majewski} S.~R.,  2012,
  \mn@doi [\apj] {10.1088/0004-637X/757/2/109}, \href
  {https://ui.adsabs.harvard.edu/abs/2012ApJ...757..109C} {757, 109}

\bibitem[\protect\citeauthoryear{{Carlberg} et~al.,}{{Carlberg}
  et~al.}{2015}]{Carlberg2015}
{Carlberg} J.~K.,  et~al., 2015, \mn@doi [\apj] {10.1088/0004-637X/802/1/7},
  \href {http://adsabs.harvard.edu/abs/2015ApJ...802....7C} {802, 7}

\bibitem[\protect\citeauthoryear{{Castelli} \& {Kurucz}}{{Castelli} \&
  {Kurucz}}{2004}]{CastelliKuruxz2004}
{Castelli} F.,  {Kurucz} R.~L.,  2004, ArXiv Astrophysics e-prints, \href
  {http://adsabs.harvard.edu/abs/2004astro.ph..5087C} {}

\bibitem[\protect\citeauthoryear{{Charbonnel} \& {Balachandran}}{{Charbonnel}
  \& {Balachandran}}{2000}]{Charbonnel2000}
{Charbonnel} C.,  {Balachandran} S.~C.,  2000, \aap, \href
  {http://adsabs.harvard.edu/abs/2000A%26A...359..563C} {359, 563}

\bibitem[\protect\citeauthoryear{{Choi}, {Dotter}, {Conroy}  \& {Ting}}{{Choi}
  et~al.}{2018}]{Choi2018}
{Choi} J.,  {Dotter} A.,  {Conroy} C.,   {Ting} Y.-S.,  2018, \mn@doi [\apj]
  {10.3847/1538-4357/aac435}, \href
  {https://ui.adsabs.harvard.edu/abs/2018ApJ...860..131C} {860, 131}

\bibitem[\protect\citeauthoryear{{Cui} et~al.,}{{Cui} et~al.}{2012}]{Cui2012}
{Cui} X.-Q.,  et~al., 2012, \mn@doi [Research in Astronomy and Astrophysics]
  {10.1088/1674-4527/12/9/003}, \href
  {https://ui.adsabs.harvard.edu/\#abs/2012RAA....12.1197C} {12, 1197}

\bibitem[\protect\citeauthoryear{{De Silva} et~al.,}{{De Silva}
  et~al.}{2015}]{galah2015}
{De Silva} G.~M.,  et~al., 2015, \mn@doi [\mnras] {10.1093/mnras/stv327}, \href
  {https://ui.adsabs.harvard.edu/\#abs/2015MNRAS.449.2604D} {449, 2604}

\bibitem[\protect\citeauthoryear{{Deepak} \& {Reddy}}{{Deepak} \&
  {Reddy}}{2019}]{deepak2019}
{Deepak} {Reddy} B.~E.,  2019, \mn@doi [\mnras] {10.1093/mnras/stz128}, \href
  {http://adsabs.harvard.edu/abs/2019MNRAS.484.2000D} {484, 2000}

\bibitem[\protect\citeauthoryear{{Denissenkov}}{{Denissenkov}}{2012}]{Denissenkov2012}
{Denissenkov} P.~A.,  2012, \mn@doi [\apjl] {10.1088/2041-8205/753/1/L3}, \href
  {http://adsabs.harvard.edu/abs/2012ApJ...753L...3D} {753, L3}

\bibitem[\protect\citeauthoryear{{Denissenkov} \& {Herwig}}{{Denissenkov} \&
  {Herwig}}{2004}]{Denissenkov2004}
{Denissenkov} P.~A.,  {Herwig} F.,  2004, \mn@doi [\apj] {10.1086/422575},
  \href {http://adsabs.harvard.edu/abs/2004ApJ...612.1081D} {612, 1081}

\bibitem[\protect\citeauthoryear{Fu, Bressan, Marigo, Girardi, Montalbán, Chen
   \& Nanni}{Fu et~al.}{2018}]{Fu2018}
Fu X.,  Bressan A.,  Marigo P.,  Girardi L.,  Montalbán J.,  Chen Y.,   Nanni
  A.,  2018, \mn@doi [Monthly Notices of the Royal Astronomical Society]
  {10.1093/mnras/sty235}, 476, 496

\bibitem[\protect\citeauthoryear{{Gaia Collaboration} et~al.,}{{Gaia
  Collaboration} et~al.}{2018}]{Gaiadr2}
{Gaia Collaboration} et~al., 2018, \mn@doi [\aap]
  {10.1051/0004-6361/201833051}, \href
  {https://ui.adsabs.harvard.edu/abs/2018A%26A...616A...1G} {616, A1}

\bibitem[\protect\citeauthoryear{{Garc{\'\i}a P{\'e}rez} et~al.,}{{Garc{\'\i}a
  P{\'e}rez} et~al.}{2016}]{Garcia2016}
{Garc{\'\i}a P{\'e}rez} A.~E.,  et~al., 2016, \mn@doi [\aj]
  {10.3847/0004-6256/151/6/144}, \href
  {https://ui.adsabs.harvard.edu/abs/2016AJ....151..144G} {151, 144}

\bibitem[\protect\citeauthoryear{{Iben}}{{Iben}}{1967a}]{Iben1967a}
{Iben} Jr. I.,  1967a, \mn@doi [\apj] {10.1086/149040}, \href
  {http://adsabs.harvard.edu/abs/1967ApJ...147..624I} {147, 624}

\bibitem[\protect\citeauthoryear{{Iben}}{{Iben}}{1967b}]{Iben1967b}
{Iben} Jr. I.,  1967b, \mn@doi [\apj] {10.1086/149041}, \href
  {http://adsabs.harvard.edu/abs/1967ApJ...147..650I} {147, 650}

\bibitem[\protect\citeauthoryear{{Jofr{\'e}}, {Petrucci}, {Garc{\'\i}a}  \&
  {G{\'o}mez}}{{Jofr{\'e}} et~al.}{2015}]{Jofre2015}
{Jofr{\'e}} E.,  {Petrucci} R.,  {Garc{\'\i}a} L.,   {G{\'o}mez} M.,  2015,
  \mn@doi [\aap] {10.1051/0004-6361/201527337}, \href
  {https://ui.adsabs.harvard.edu/abs/2015A&A...584L...3J} {584, L3}

\bibitem[\protect\citeauthoryear{{Kirby}, {Guhathakurta}, {Zhang}, {Hong},
  {Guo}, {Guo}, {Cohen}  \& {Cunha}}{{Kirby} et~al.}{2016}]{Kirby2016}
{Kirby} E.~N.,  {Guhathakurta} P.,  {Zhang} A.~J.,  {Hong} J.,  {Guo} M.,
  {Guo} R.,  {Cohen} J.~G.,   {Cunha} K.,  2016, \mn@doi [\apj]
  {10.3847/0004-637X/819/2/135}, \href
  {http://adsabs.harvard.edu/abs/2016ApJ...819..135K} {819, 135}

\bibitem[\protect\citeauthoryear{{Kjeldsen} \& {Bedding}}{{Kjeldsen} \&
  {Bedding}}{1995}]{Bedding1995}
{Kjeldsen} H.,  {Bedding} T.~R.,  1995, \aap, \href
  {http://adsabs.harvard.edu/abs/1995A%26A...293...87K} {293, 87}

\bibitem[\protect\citeauthoryear{{Kumar}, {Reddy}  \& {Lambert}}{{Kumar}
  et~al.}{2011}]{Kumar2011a}
{Kumar} Y.~B.,  {Reddy} B.~E.,   {Lambert} D.~L.,  2011, \mn@doi [\apjl]
  {10.1088/2041-8205/730/1/L12}, \href
  {http://adsabs.harvard.edu/abs/2011ApJ...730L..12K} {730, L12}

\bibitem[\protect\citeauthoryear{{Lagarde}, {Decressin}, {Charbonnel},
  {Eggenberger}, {Ekstr{\"o}m}  \& {Palacios}}{{Lagarde}
  et~al.}{2012}]{Lagarde2012}
{Lagarde} N.,  {Decressin} T.,  {Charbonnel} C.,  {Eggenberger} P.,
  {Ekstr{\"o}m} S.,   {Palacios} A.,  2012, \mn@doi [\aap]
  {10.1051/0004-6361/201118331}, \href
  {http://adsabs.harvard.edu/abs/2012A%26A...543A.108L} {543, A108}

\bibitem[\protect\citeauthoryear{{Lambert} \& {Reddy}}{{Lambert} \&
  {Reddy}}{2004}]{Lambert2004}
{Lambert} D.~L.,  {Reddy} B.~E.,  2004, \mn@doi [\mnras]
  {10.1111/j.1365-2966.2004.07557.x}, \href
  {http://adsabs.harvard.edu/abs/2004MNRAS.349..757L} {349, 757}

\bibitem[\protect\citeauthoryear{{Lind}, {Primas}, {Charbonnel}, {Grundahl}  \&
  {Asplund}}{{Lind} et~al.}{2009}]{Lind2009b}
{Lind} K.,  {Primas} F.,  {Charbonnel} C.,  {Grundahl} F.,   {Asplund} M.,
  2009, \mn@doi [\aap] {10.1051/0004-6361/200912524}, \href
  {http://adsabs.harvard.edu/abs/2009A%26A...503..545L} {503, 545}

\bibitem[\protect\citeauthoryear{{Liu}, {Tan}, {Wang}, {Zhao}, {Sato}, {Takeda}
   \& {Li}}{{Liu} et~al.}{2014}]{Liu2014}
{Liu} Y.~J.,  {Tan} K.~F.,  {Wang} L.,  {Zhao} G.,  {Sato} B.,  {Takeda} Y.,
  {Li} H.~N.,  2014, \mn@doi [\apj] {10.1088/0004-637X/785/2/94}, \href
  {http://adsabs.harvard.edu/abs/2014ApJ...785...94L} {785, 94}

\bibitem[\protect\citeauthoryear{{Majewski} et~al.,}{{Majewski}
  et~al.}{2017}]{Majewski2017}
{Majewski} S.~R.,  et~al., 2017, \mn@doi [\aj] {10.3847/1538-3881/aa784d},
  \href {https://ui.adsabs.harvard.edu/abs/2017AJ....154...94M} {154, 94}

\bibitem[\protect\citeauthoryear{{Matsuno}, {Yong}, {Aoki}  \&
  {Ishigaki}}{{Matsuno} et~al.}{2018}]{Matsuno2018}
{Matsuno} T.,  {Yong} D.,  {Aoki} W.,   {Ishigaki} M.~N.,  2018, \mn@doi [\apj]
  {10.3847/1538-4357/aac019}, \href
  {https://ui.adsabs.harvard.edu/abs/2018ApJ...860...49M} {860, 49}

\bibitem[\protect\citeauthoryear{{Mosser} et~al.,}{{Mosser}
  et~al.}{2014}]{Mosser2014}
{Mosser} B.,  et~al., 2014, \mn@doi [\aap] {10.1051/0004-6361/201425039}, \href
  {https://ui.adsabs.harvard.edu/abs/2014A&A...572L...5M} {572, L5}

\bibitem[\protect\citeauthoryear{{Palacios}, {Charbonnel}  \&
  {Forestini}}{{Palacios} et~al.}{2001}]{Palacios2001}
{Palacios} A.,  {Charbonnel} C.,   {Forestini} M.,  2001, \mn@doi [\aap]
  {10.1051/0004-6361:20010903}, \href
  {http://adsabs.harvard.edu/abs/2001A%26A...375L...9P} {375, L9}

\bibitem[\protect\citeauthoryear{{Ram{\'{\i}}rez} \& {Allende
  Prieto}}{{Ram{\'{\i}}rez} \& {Allende Prieto}}{2011}]{Ramirez2011}
{Ram{\'{\i}}rez} I.,  {Allende Prieto} C.,  2011, \mn@doi [\apj]
  {10.1088/0004-637X/743/2/135}, \href
  {http://adsabs.harvard.edu/abs/2011ApJ...743..135R} {743, 135}

\bibitem[\protect\citeauthoryear{{Reddy}, {Lambert}, {Laws}, {Gonzalez}  \&
  {Covey}}{{Reddy} et~al.}{2002}]{Reddy2002}
{Reddy} B.~E.,  {Lambert} D.~L.,  {Laws} C.,  {Gonzalez} G.,   {Covey} K.,
  2002, \mn@doi [\mnras] {10.1046/j.1365-8711.2002.05682.x}, \href
  {http://adsabs.harvard.edu/abs/2002MNRAS.335.1005R} {335, 1005}

\bibitem[\protect\citeauthoryear{{Reddy}, {Tomkin}, {Lambert}  \& {Allende
  Prieto}}{{Reddy} et~al.}{2003}]{Reddy2003}
{Reddy} B.~E.,  {Tomkin} J.,  {Lambert} D.~L.,   {Allende Prieto} C.,  2003,
  \mn@doi [\mnras] {10.1046/j.1365-8711.2003.06305.x}, \href
  {http://adsabs.harvard.edu/abs/2003MNRAS.340..304R} {340, 304}

\bibitem[\protect\citeauthoryear{{Siess} \& {Livio}}{{Siess} \&
  {Livio}}{1999}]{Siess1999}
{Siess} L.,  {Livio} M.,  1999, \mn@doi [\mnras]
  {10.1046/j.1365-8711.1999.02784.x}, \href
  {http://adsabs.harvard.edu/abs/1999MNRAS.308.1133S} {308, 1133}

\bibitem[\protect\citeauthoryear{{Silva Aguirre} et~al.,}{{Silva Aguirre}
  et~al.}{2014}]{SilvaAguirre2014}
{Silva Aguirre} V.,  et~al., 2014, \mn@doi [\apjl]
  {10.1088/2041-8205/784/1/L16}, \href
  {http://adsabs.harvard.edu/abs/2014ApJ...784L..16S} {784, L16}

\bibitem[\protect\citeauthoryear{{Singh}, {Reddy}, {Bharat Kumar}  \&
  {Antia}}{{Singh} et~al.}{2019}]{Singh2019l}
{Singh} R.,  {Reddy} B.~E.,  {Bharat Kumar} Y.,   {Antia} H.~M.,  2019, \mn@doi
  [\apjl] {10.3847/2041-8213/ab2599}, \href
  {https://ui.adsabs.harvard.edu/abs/2019ApJ...878L..21S} {878, L21}

\bibitem[\protect\citeauthoryear{{Smiljanic} et~al.,}{{Smiljanic}
  et~al.}{2018}]{Smiljanic2018}
{Smiljanic} R.,  et~al., 2018, \mn@doi [\aap] {10.1051/0004-6361/201833027},
  \href {https://ui.adsabs.harvard.edu/abs/2018A&A...617A...4S} {617, A4}

\bibitem[\protect\citeauthoryear{{Sneden}}{{Sneden}}{1973}]{Sneden1973}
{Sneden} C.~A.,  1973, PhD thesis, THE UNIVERSITY OF TEXAS AT AUSTIN.

\bibitem[\protect\citeauthoryear{{Takeda} \& {Tajitsu}}{{Takeda} \&
  {Tajitsu}}{2017}]{Takeda2017}
{Takeda} Y.,  {Tajitsu} A.,  2017, \mn@doi [\pasj] {10.1093/pasj/psx057}, \href
  {https://ui.adsabs.harvard.edu/abs/2017PASJ...69...74T} {69, 74}

\bibitem[\protect\citeauthoryear{{Tayar} et~al.,}{{Tayar}
  et~al.}{2017}]{Tayar2017}
{Tayar} J.,  et~al., 2017, \mn@doi [\apj] {10.3847/1538-4357/aa6a1e}, \href
  {https://ui.adsabs.harvard.edu/abs/2017ApJ...840...17T} {840, 17}

\bibitem[\protect\citeauthoryear{{Wallerstein} \& {Sneden}}{{Wallerstein} \&
  {Sneden}}{1982}]{Wallerstein1982}
{Wallerstein} G.,  {Sneden} C.,  1982, \mn@doi [\apj] {10.1086/159859}, \href
  {https://ui.adsabs.harvard.edu/abs/1982ApJ...255..577W} {255, 577}

\bibitem[\protect\citeauthoryear{{Yan} et~al.,}{{Yan} et~al.}{2018}]{yan2018}
{Yan} H.-L.,  et~al., 2018, \mn@doi [Nature Astronomy]
  {10.1038/s41550-018-0544-7}, \href
  {https://ui.adsabs.harvard.edu/abs/2018NatAs...2..790Y} {2, 790}

\bibitem[\protect\citeauthoryear{Yong et~al.,}{Yong et~al.}{2016}]{Yong2016}
Yong D.,  et~al., 2016, \mn@doi [\mnras] {10.1093/mnras/stw676}, 459, 487

\bibitem[\protect\citeauthoryear{{Yu}, {Huber}, {Bedding}, {Stello}, {Hon},
  {Murphy}  \& {Khanna}}{{Yu} et~al.}{2018}]{Yu2018}
{Yu} J.,  {Huber} D.,  {Bedding} T.~R.,  {Stello} D.,  {Hon} M.,  {Murphy}
  S.~J.,   {Khanna} S.,  2018, \mn@doi [\apjs] {10.3847/1538-4365/aaaf74},
  \href {https://ui.adsabs.harvard.edu/abs/2018ApJS..236...42Y} {236, 42}

\bibitem[\protect\citeauthoryear{{da Silva} et~al.,}{{da Silva}
  et~al.}{2006}]{desilva2006}
{da Silva} L.,  et~al., 2006, \mn@doi [\aap] {10.1051/0004-6361:20065105},
  \href {https://ui.adsabs.harvard.edu/abs/2006A%26A...458..609D} {458, 609}

\bibitem[\protect\citeauthoryear{{de La Reza}, {Drake}  \& {da Silva}}{{de La
  Reza} et~al.}{1996}]{Delareza1996}
{de La Reza} R.,  {Drake} N.~A.,   {da Silva} L.,  1996, \mn@doi [\apjl]
  {10.1086/309874}, \href
  {https://ui.adsabs.harvard.edu/abs/1996ApJ...456L.115D} {456, L115}

\makeatother
\end{thebibliography}

\bsp	
\label{lastpage}
\end{document}